\documentclass[
 twocolumn,superscriptaddress,aps,amsmath,amssymb,floatfix]{revtex4-2}
\usepackage{comment}
\usepackage{graphicx}% Include figure files
\usepackage{dcolumn}% Align table columns on decimal point
\usepackage{bm}% bold math
\usepackage{dsfont}
\usepackage[breaklinks=true,colorlinks,citecolor=blue,linkcolor=blue,urlcolor=blue]{hyperref}% add hypertext capabilities

\usepackage{physics}
\usepackage{mleftright}
\usepackage{bbm}

\newcommand{\be}{\begin{equation}}
\newcommand{\ee}{\end{equation}}
\newcommand{\bea}{\begin{eqnarray}}
\newcommand{\eea}{\end{eqnarray}}

\newcommand{\figref}[1]{\mbox{Fig.~\ref{#1}}}

\renewcommand{\eqref}[1]{\mbox{Eq.~(\ref{#1})}}

\newcommand{\figpanel}[2]{Fig.~\hyperref[#1]{\ref*{#1}(#2)}}

\begin{document}

\title{Optomechanical Two-Photon Hopping}

\author{Enrico Russo}
\email{enricorussoxvi@gmail.com}
\affiliation{Dipartimento di Scienze Matematiche e Informatiche, Scienze Fisiche e  Scienze della Terra, Universit\`{a} di Messina, I-98166 Messina, Italy}

\author{Alberto Mercurio}
%\email{alberto.mercurio@unime.it}
\affiliation{Dipartimento di Scienze Matematiche e Informatiche, Scienze Fisiche e  Scienze della Terra, Universit\`{a} di Messina, I-98166 Messina, Italy}

\author{Fabio Mauceri}
\affiliation{Dipartimento di Scienze Matematiche e Informatiche, Scienze Fisiche e  Scienze della Terra, Universit\`{a} di Messina, I-98166 Messina, Italy}

\author{Rosario Lo Franco}
\affiliation{Dipartimento di Ingegneria, Universit\`{a} degli Studi di Palermo, Viale delle Scienze, 90128 Palermo, Italy}

\author{Franco Nori}
\affiliation{Theoretical Quantum Physics Laboratory, RIKEN, Wako-shi, Saitama 351-0198, Japan}
\affiliation{RIKEN Center for Quantum Computing (RQC), Wako-shi, Saitama 351-0198, Japan}
\affiliation{Physics Department, The University of Michigan, Ann Arbor, Michigan 48109-1040, USA}

\author{Salvatore Savasta}
\affiliation{Dipartimento di Scienze Matematiche e Informatiche, Scienze Fisiche e Scienze della Terra, Universit\`{a} di Messina, I-98166 Messina, Italy}
\affiliation{Theoretical Quantum Physics Laboratory, RIKEN, Wako-shi, Saitama 351-0198, Japan}

\author{Vincenzo Macr\`{i}}
\email{vincenzo.macri@riken.jp}
\affiliation{Theoretical Quantum Physics Laboratory, RIKEN, Wako-shi, Saitama 351-0198, Japan}

\affiliation{Dipartimento di Ingegneria, Universit\`{a} degli Studi di Palermo, Viale delle Scienze, 90128 Palermo, Italy}

\date{\today}% It is always \today, today,

\begin{abstract}
%The hopping mechanism plays a key role in collective phenomena emerging in many-body physics. The ability to create and control systems that display this feature is important for next generation quantum technologies. Here we propose an optomechanical model, with experimentally attainable range of parameters, capable of displaying photon-pair hopping between two electromagnetic resonators. The physics of two cavities separated by a vibrating two-sided perfect mirror is analysed starting from its canonical quantisation. Although in the classical realm, the vibrating mirror perfectly separates the two sides of the cavity, in the quantum regime these two sides can interact. Such a system opens the possibility to investigate a new mechanism of photon-pair propagation in optomechanical lattices within different experimental platforms. 
The hopping mechanism plays a key role in collective phenomena emerging in many-body physics. The ability to create and control systems that display this feature is important for next generation quantum technologies. Here we study two cavities separated by a vibrating two-sided perfect mirror and show that, within currently available experimental parameters, this system displays photon-pair hopping between the two electromagnetic resonators. In particular, the two-photon hopping is not due to tunneling, but rather to higher-order resonant processes.
Starting from the classical problem, where the vibrating mirror perfectly separates the two sides of the cavity, we quantize the system and then the two sides can interact. This opens the possibility to investigate a new mechanism of photon-pair propagation in optomechanical lattices.
\end{abstract}

%Here we study two cavities separated by a vibrating two-sided perfect mirror and show that, within to-day experimental parameters, this system displays photon-pair hopping between the two electromagnetic resonators.

%, within to-day experimental parameters, 

\maketitle

The mastery of manipulating quantum mechanical systems by means of radiation pressure has opened the door to fundamental tests of quantum theory \cite{Marshall2003,Gavartin2012}, to precision measurements \cite{Teufel2009,Krause2012,Carlesso2021} and to novel quantum technologies \cite{Anetsberger2009,Verlot2009,Barzanjeh2022}. For instance, laser cooling techniques \cite{Schliesser2009,Groblacher2009,Teufel2011} allow to observe quantized vibrational modes of macroscopic objects and even the possibility to reach their ground-state \cite{Wilson2007,Chan2011,Ojanen2014}. This has paved the way to the realisation of entangled macroscopic states and, in turn, new ways to process and store quantum information \cite{Stannigel2012,Garziano2015pra,meystre2013,Macri2016}. Notably, with these techniques optomechanical crystals \cite{Eichenfield2009,EEichenfield2009,zhang2018} can be scaled to form optomechanical arrays where, using hopping mechanisms, applications for quantum information processing have been proposed \cite{chang2011,schmidt2012}.

Cavity-optomechanics, in particular, lies at the crossroad of wide research lines that are currently under active investigation. 
In experiments \cite{Barzanjeh2022, meystre2013, Aspelmeyer2014, khalili2016}, only radiation pressure effects have been considered, as the cavity frequencies far outweigh the mirror ones. On the other hand, ultra-high-frequency mechanical oscillators \cite{Connell2010,rouxinol2016} coupled to microwave ones offer the potential to observe, for instance, dynamical Casimir effects \cite{Moore1970,Norirmp,Johansson2009,Johansson2010,Wilson2011,Dodonov2020}.
The case of %a single ultra-high-frequency movable mirror 
one such mirror interacting with a single cavity mode was first considered in Ref. \cite{Macri2018}, and this study was later extended to include an incoherent excitation of the mirror \cite{Settineri2018,Settineri2019,Fong2019}. In the same setup, back-reaction and dissipation effects have also been studied \cite{Butera2019,ferreri2022}. Finally, the case of a cavity with two moving walls was addressed  \cite{DiStefano2019,Butera22}; in this case, the cavity field mediates an effective interaction between the two mirrors leading to a phonon hopping.
\begin{figure}[b]
    \centering
    \includegraphics[width = \linewidth]{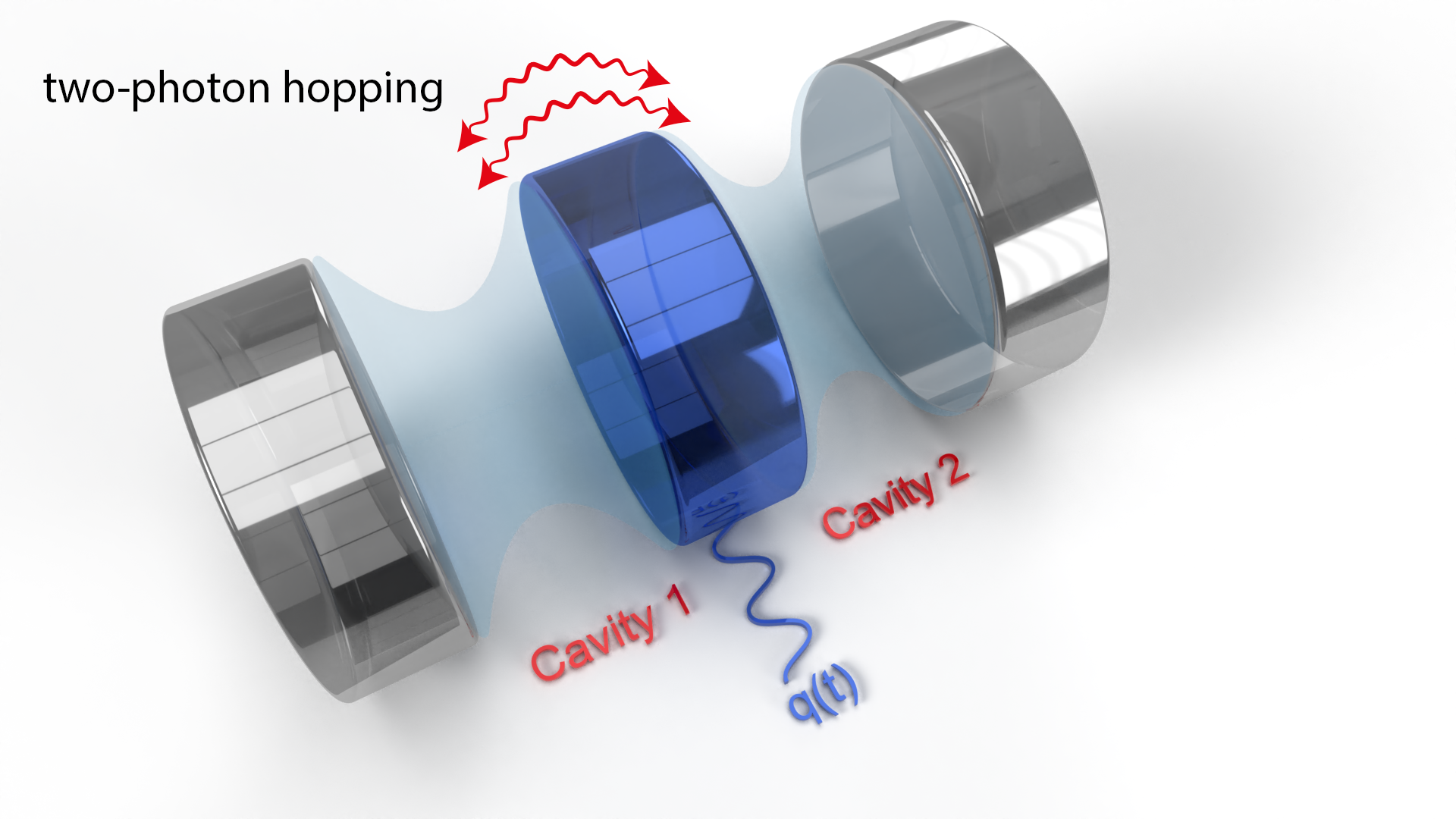}
    \caption{Proposal sketch. Two non-interacting electromagnetic cavities separated by a movable two-sided perfect mirror.}
    \label{fig:sketch}
\end{figure}
%a cavity with two moving walls has been considered in Ref. \cite{DiStefano2019,Butera22}. In the latter case, cavity field mediates an effective interaction between the two mirrors leading to a phonon hopping.
%When describing optomechanical experiments to date \cite{Aspelmeyer2014,khalili2016}, due to the low-frequency mechanical oscillator compared to the cavity one, Hamiltonian terms which convert mechanical energy into electromagnetic one are neglected \cite{Law1995}. It turns out that, for ultra-high-frequency mechanical oscillators, it shall not be so. By considering the mechanical-electromagnetic energy conversion terms, several theoretical work have been proposed, for instance, the possibility to observe the dynamical Casimir effect between one mirror and a resonator \cite{Macri2018} and studying dissipation and decoherence effects on the motion of moving objects \cite{Butera2019,ferreri2022}, even when the source of the mechanical excitation is a thermal bath \cite{Settineri2018}. Moreover, mechanical excitations transfer, when consider an optomechanical system consisting of two vibrating mirrors constituting an optical resonator \cite{DiStefano2019}, can be predicted.

A suitable platform to experimentally reproduce these predictions is circuit optomechanics. %with ultra-high-frequency (5-6 GHz) mechanical oscillators. 
In fact, the addition of artificial atoms in a superconducting microwave setup strengthens the coupling with the mechanical resonator \cite{rouxinol2016,Heikkil2014,Pirkkalainen2015}, and introducing high-frequency mirrors makes it a very promising setup.
A valuable alternative would be to use a quantum simulator \cite{Johansson2014,Kim2015} where two \emph{LC} circuits replace the two cavities, and a superconducting quantum interference device (SQUID) is deployed instead of the high-frequency vibrating mirror. 
%The interaction with a superconducting microwave resonator, through artificial atoms \cite{rouxinol2016,Heikkil2014,Pirkkalainen2015}, should allow the observation of the effects predicted here. 
%In particular, a combination of circuit optomechanical schemes with, already demonstrated, ultra-high frequency mechanical resonators represents a very promising setup. %\cite{Kim2015}. 

%Inspired by these experimental achievements, we suggest here a feasible system that serves as a building block for future investigations. 
The availability of these experimental platforms led us to design a system that, under certain resonance conditions,
%Its phenomenology is rich and we have found a peculiar dynamics that 
allows for a simultaneous hopping of photon-pairs. The system consists of two non-interacting electromagnetic resonators separated by a movable two-sided perfect mirror. The vibrational modes of the mirror act as a mediator between the two resonators, making the photon-pair hopping possible. The vibrating mirror separates both sides of the cavity at the classical level, but not quantum mechanically.
%What is separated according to classical physics turns out to be coupled according to quantum mechanics. 
Our Hamiltonian is obtained quantizing the classical problem, generalizing the results in Ref.~\cite{Law1995}. It accounts also for generic equilibrium positions of the mirror even though, in what follows, we consider only the symmetric case. 

Similar setups have been studied, for instance, in Ref. \cite{montalbano2022} where the authors analyzed the dressing of the ground state and the correlation functions between the two separated regions, and in Ref. \cite{Cheung2011} where the two resonators are separated by a dielectric. In our treatment the two-photon hopping mechanism appears as a spontaneous coherent process in a second-order effective dynamics. Note that the optomechanical hopping described here does not involve photon tunneling, which is the usual photon hopping mechanism studied elsewhere. 
%Our setup has also the virtue of converting classic incoming light into non-classical squeezed light.
Our interest in these hopping effects stems from the possibility to envision optomechanical lattices, with unit cells as in \figref{fig:sketch}, and to study their thermodynamic and information properties. Thus, extended optomechanical lattices would display an interesting interplay between the Casimir photon-pairs creation and the lattice inter-site hopping.

%\vspace{4mm}
%\noindent \textbf{Results}

\section*{Results}

\begin{figure}[t]
    \centering
    \includegraphics[width = \linewidth]{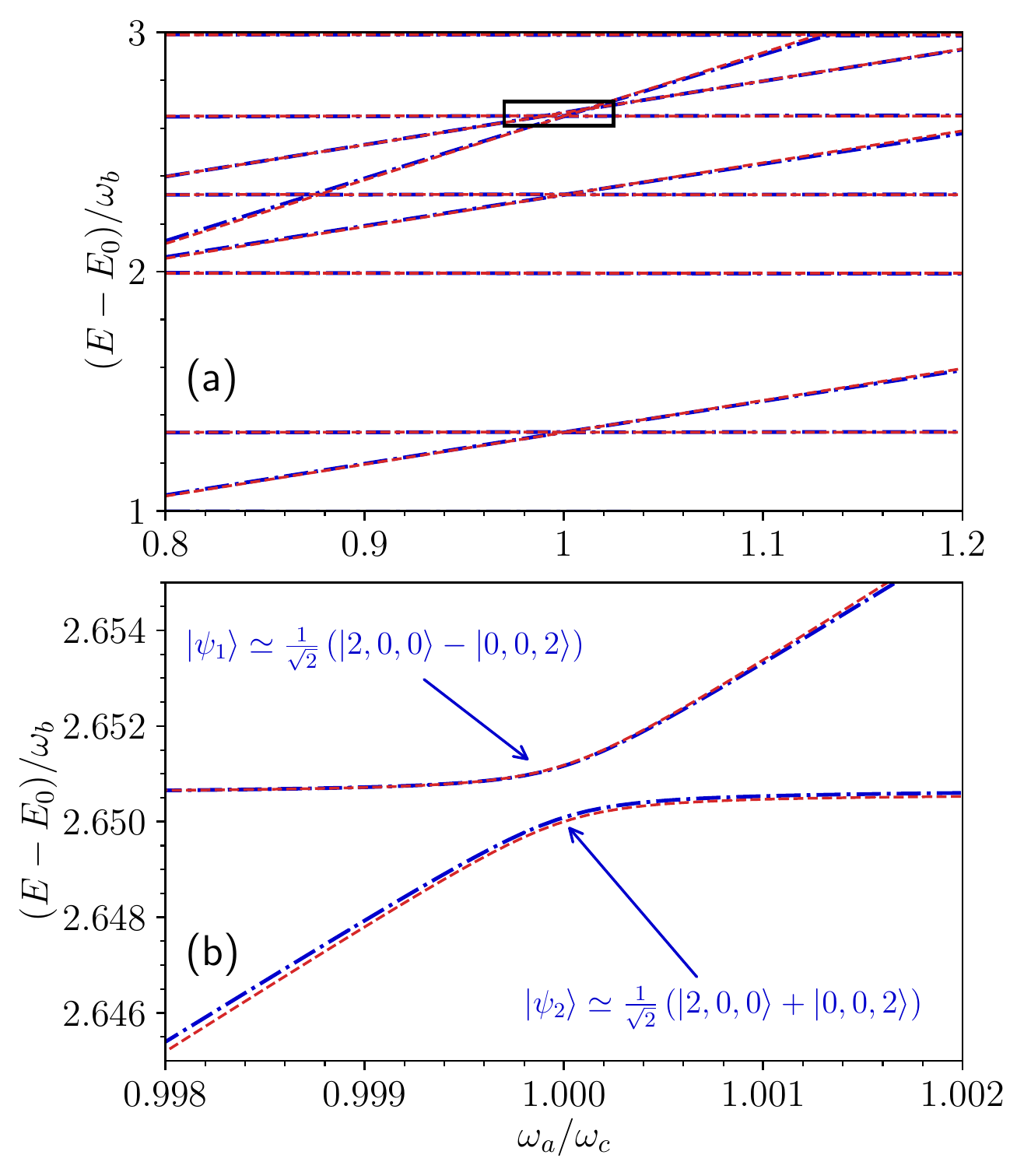}
    \caption{(a) The lowest energy levels of the system Hamiltonian versus the ratio between the two cavity frequencies. For a coupling $g= 0.06\,\omega_b$, the position of the avoid level crossing is contained in the black rectangular. (b) An enlarged view of the latter is given. The presence of the labels stress the hybridisation of the two states $\ket{2,0,0}$ and $\ket{0,0,2}$. The frequency mirror was conveniently set as $\omega_b = 3/4\, \omega_c$.}
    \label{hopping_position}
\end{figure}

\noindent \textbf{The quantum model.} Consider two non-interacting electromagnetic cavities separated by a vibrating two-sided perfect mirror as sketched in \figref{fig:sketch}. Following Ref.~\cite{Law1995} we quantized (see Methods) the classical system obtaining the Hamiltonian ($\hbar=1$)
\begin{align}\label{Hamiltonian}
    \hat{H}  &=  \omega_a  \hat{a}^\dagger  \hat{a} + \omega_b  \hat{b}^\dagger  \hat{b} + \omega_c  \hat{c}^\dagger  \hat{c} \\
  &+ \frac{g}{2}\left[  ( \hat{c}+ \hat{c}^\dagger)^2- \bigg(\frac{\omega_a}{\omega_c} \bigg)^2 ( \hat{a}+ \hat{a}^\dagger)^2  \right]( \hat{b}+ \hat{b}^\dagger) \nonumber \, .
\end{align}
Here, $\hat b $ $(\hat b^\dag)$ is the creation (annihilation) operator of the moving mirror, $\hat a $  $  (\hat a^\dag)$ and  $\hat c $  $  (\hat c^\dag)$ are the creation (annihilation) operators of the left and right cavity, respectively. The parameters $\omega_a$, $\omega_b$ and $\omega_c$ are the corresponding bare energies of the three boson modes. The coupling strength $g=\omega_c^2x_{\rm zpf}/\pi $ depends both on the zero-point-fluctuation amplitude of the mirror $x_{\rm zpf}$, and on the bare energy of a cavity $\omega_c$, taken for convenience as the right one. The weight $\omega_a^2/\omega_c^2$ accounts for asymmetrical configurations. The linear approximation implicit in \eqref{Hamiltonian} does not lead to instabilities of the ground state as long as $g \omega_a<\omega_c^2 $, i.e., $\omega_a x_{\rm zpf} < \pi$. 
%The minus sign in \eqref{Hamiltonian} leads to instabilities of the ground state unless $g \omega_a<\omega_c^2 $, i.e., $\omega_a x_{\rm zpf} < \pi$.
%With this conditions satisfied, the system can reach the USC regime within the linear approximation.
%In order to describe the two photon-pairs hopping dynamics, we solve numerically the spectrum of the Hamiltonian and thus locate avoid-level crossings under the resonance $\omega_a = \omega_c$. For what concerns the mirror frequencies,
The sought-after hopping mechanism occurs at the resonance $\omega_a=\omega_c$. 
We consider the case when the bare frequency of the mirror is lower than the cavity frequency.
%For what concerns the mirror, looking at the experimental attainable setup in the state-of-art \cite{Connell2010,Fong2019}, it would be preferable to have the resonance frequency of the mirror lower than the cavity one. 
%Besides, the aforementioned circuital setups have the advantage to control the photon-phonon coupling by means of an additional superconducting qubit, dispersively coupled to the cavity \cite{rouxinol2016}. 
%Indeed, coupling such a mechanical oscillator to a superconducting qubit, which can be dispersive coupled to a cavity resonator, allow to quantum control over a macroscopic mechanical system as it has been demonstrated \cite{Connell2010,rouxinol2016}. Therefore, the effects described here can be experimentally demonstrated with circuit-optomechanical systems, using ultra-high-frequency mechanical micro- or nano-resonators in the GHz spectral range \cite{Connell2010}, 
This choice of parameters identifies a set of avoided-level crossings in the Hamiltonian spectrum, and thus a particular closed sub-dynamics,
as can be seen from \figref{hopping_position}, %where the frequency mirror was conveniently set as $\omega_b = 3/4\, \omega_c$.

%for this reason, our results are carried on for $\omega_b=3/4 \omega_a$.
%Under this choice of parameters a regions where our specific avoided level crossing appears is shown in \figref{hopping_position}.
\begin{comment}
In order to describe the two photon-pairs hopping dynamics, we solve numerically the spectrum of the Hamiltonian and thus locate avoid-level crossings under the resonance $\omega_a = \omega_c$.
Looking at the experimental attainable setup in the state-of-art \cite{Connell2010,rouxinol2016,Fong2019} it would be preferable to have a frequency of the mirror lower than the cavity one. Indeed, coupling such a mechanical oscillator to a superconducting qubit, which can be dispersive coupled to a cavity resonator, allow to quantum control over a macroscopic mechanical system as it has been demonstrated \cite{Connell2010,rouxinol2016}. Therefore, the effects described here can be experimentally demonstrated with circuit-optomechanical systems, using ultra-high-frequency mechanical micro- or nano-resonators in the GHz spectral range \cite{Connell2010}, for this reasoner, our results are carried on for $\omega_b=3/4 \omega_a$. 
Under this choice of parameters a regions where our specific avoided level crossing appears is shown in \figpanel{hopping_position}{b,c}. It shows a gap which is a trademark of an hybridisation of the two states $\ket{\psi_1}$ and $\ket{\psi_2}$, eigenstates of the full Hamiltonian \eqref{Hamiltonian}.
\end{comment}

 Figure~\ref{hopping_position}{\textcolor{blue}{(a)}} shows the lowest energy levels obtained by numerically diagonalising the full Hamiltonian \eqref{Hamiltonian} (blue dashdotted curves), while \figpanel{hopping_position}{b} is an enlarged view of the avoided-level crossing inside the black rectangle. The gap is a trademark of the hybridization of the two states $\ket{\psi_1}$ and $\ket{\psi_2}$, eigenstates of the full Hamiltonian \eqref{Hamiltonian}. A local effective description (red dashed curves) is possible through the generalized James' effective approach \cite{Shao2017} (see Methods), with resonance conditions $\omega_c = \omega_a$ 
\bea
\label{Hopping_Second-order}
\hat H_{\rm eff}^{(2)} &=& \hat{H}_{\rm shift}^{(2)} + \hat{H}_{\rm hop}^{(2)} \, , \nonumber \\
\hat{H}_{\rm shift}^{(2)} &=&  \left [\omega_a + \frac{g^2(4\omega_a+\omega_b)}{8\omega_a^2-2\omega^2_b}\right]( \hat{c}^\dagger \hat{c} + \hat{a}^\dagger \hat{a}) \nonumber \\
&&+ \frac{g^2(3\omega_b^2- 8 \omega^2_a)}{(8\omega_a^2-2\omega_b^2)\omega_b} \left[ (\hat{c}^\dagger \hat{c})^2 + (\hat{a}^\dagger \hat{a})^2 \right] \nonumber \\
&&+ \left[ \omega_b +  \frac{4 g^2 \omega_a}{4\omega_a^2-\omega^2_b} (\hat{a}^\dagger \hat{a} + \hat{c}^\dagger \hat{c} + \mathds{1})\right] \hat{b}^\dagger \hat{b} \nonumber \\
&&+ \frac{2g^2}{\omega_b} \hat{a}^\dagger \hat{a} \hat{c}^\dagger \hat{c} + \frac{g^2}{2\omega_a-\omega_b} \mathds{1} \nonumber \, , \\
\hat{H}_{\rm hop}^{(2)} &=&  -\frac{g^2 \omega_b}{8\omega_a^2-2\omega^2_b} (\hat{a}^2 \hat{c}{ ^\dagger}^2 + \hat{a}{^\dagger}^2 \hat{c}^2 ) \, .
\eea
The first term, $\hat{H}_{\rm shift}^{(2)}$, contains the bare Hamiltonians and both cross- and self-Kerr non-linearities. The second term, $\hat{H}_{ \rm hop}^{(2)}$ is the one responsible for the two-photon hopping. 
Since $[\hat{a}^\dagger \hat{a},\hat{H}_{\rm shift}^{(2)}] = [\hat{b}^\dagger \hat{b},\hat{H}_{\rm shift}^{(2)}] = [\hat{c}^\dagger \hat{c},\hat{H}_{\rm shift}^{(2)}] = 0 $ we can still choose as an unperturbed base the states $\ket{n_a, n_b, n_c}$, where $n_a$ ($n_c$) is the number of photon in the left (right) cavity, and $n_b$ the number of phonons in between; all of these three are considered with shifted energies due to interaction with the fields.\\
%and the number operators, $\hat{a}^\dagger \hat{a}$ $\hat{b}^\dagger \hat{b}$ and $\hat{c}^\dagger \hat{c}$, commute with this shift Hamiltonian; thus we can still refer to the ket $\ket{n_a,n_b,n_c}$ as describing $n_a$, $n_b$  and $n_c$ bosons in the left cavity, movable mirror and the right cavity, respectively. The energy of the state $\ket{n_a, n_b, n_c}$ is shifted compared to the common use in the literature.
%To conclude, the $\hat{H}_I^{(2)}$ term describes the photon-pair hopping process and will be treated perturbatively.
%We can still legitimately adopt the notation $\ket{n_a,n_b,n_c}$ to describe $n_a$, $n_b$ and $n_c$ bosons in the left cavity, movable mirror and the right cavity, respectively, because the number operators, $\hat{a}^\dagger \hat{a}$ $\hat{b}^\dagger \hat{b}$ and $\hat{c}^\dagger \hat{c}$, commute with the shift Hamiltonian. We can thus use these states to describe the photon-pair hopping dynamics stemming from $\hat{H}_I^{(2)}$. The only difference with its common use in literature is that now their energy is shifted.
%In particular, $\hat{H}_{\rm shift}^{(2)}$ shifts the  bare energy spectrum, while $\hat{H}_I^{(2)}$ describes the hopping process of a photon-pair.
%Since the effective Hamiltonian contains  energy-shift terms, we denote as {\emph{dressed}} states the ket $\ket{n_a,n_b,n_c}$ which has $n_a$, $n_b$  and $n_c$ number of occupation of the left cavity, movable mirror and the right cavity, respectively.

\noindent \textbf{Analytical aspects.} 
The two states $\ket{\psi_{1,2}} = \left(\ket{2,0,0} \pm \ket{0,0,2} \right)/\sqrt{2}$ are eigenstates of the full (effective) Hamiltonian.
To have a simple analytical description, we limit our analysis to the
subspace spanned by $\{ \ket{2,0,0}, \ket{0,0,2} \}$ around the avoided-level crossing.
If we initialise the system in either $\ket{2,0,0}$ or $\ket{0,0,2}$, we witness a coherent oscillatory dynamics between the two maximally entangled photon-pair states.
Neglecting  dressing energy shifts, which have been reabsorbed by an appropriate choice of the coefficients,
the effective interaction Hamiltonian  $\hat H_{\rm hop}^{(2)}$ in \eqref{Hopping_Second-order} can be used to solve the stochastic evolution of the system wave function (see Methods). By projecting the time-evolution operator $\hat U(t) = \exp ( - i \mathcal{\hat H} t )$ onto the $2D$ subspace $\{ \ket{2,0,0}, \ket{0,0,2} \}$, with 
\be 
\mathcal{\hat H}= \hat H_{\rm hop}^{(2)}-i(\gamma_a \hat{a}^{\dag}\hat{a}+\gamma_b \hat{b}^{\dag}\hat{b}+\gamma_c \hat{c}^{\dag}\hat{c})/2\;,
\ee 
in the interaction picture  we obtain 
\bea
\label{evolution_operator} 
\hat U(t) &=& e^{-2\gamma t} 
\left [\cos\left( \tilde{g} t \right) \left ( \ket{2,0,0}\bra{2,0,0}+ \ket{0,0,2}\bra{0,0,2}\right) \right. \nonumber \\
&&- \left. i \sin\left( \tilde{g} t \right) \left(\ket{2,0,0}\bra{0,0,2}+ \ket{0,0,2}\bra{2,0,0}\right) \right] \, , \nonumber \\
\eea
where we choose $\gamma=\gamma_a=\gamma_c$ and $\tilde{g}=  g^2\omega_b/2(4\omega_a^2-\omega^2_b)$. If we initialize the system in the state $\ket{2,0,0}$, its evolution at time $t$, before a quantum jump takes place, is
	\begin{equation}
	\label{evolution_state} 
	\ket {\psi (t)}=e^{-2\gamma t} 
	\left [\cos\left( \tilde{g} t \right)\ket{2,0,0}-i \sin\left( \tilde{g} t \right) \ket{0,0,2}	\right ] \, .
	\end{equation}
	By appropriately renormalizing the wave function, we obtain the mean photon number for the left and right cavities and for the mechanical resonator
	\bea
	\label{evolution_Mean_photon_number} 
	\langle \hat{a}^{\dagger} \hat{a} \rangle &=& 
	2\cos^2\left(\tilde{g} t \right), \nonumber\\ 
	\langle \hat{b}^{\dagger} \hat{b} \rangle &=& 0 \,, \nonumber\\
	\langle \hat{c}^{\dagger} \hat{c} \rangle  &=& 2 \sin^2 \left( \tilde{g} t \right) \, .
	\eea
The expectation values on a single quantum trajectory for a generic operator $\hat{O}$ is denoted as $ \langle \hat{O}(t) \rangle$, while average quantities obtained over an ideally infinite number of quantum trajectories are indicated as $ \overline{\langle \hat{O}(t) \rangle} $. \\
 \begin{figure}[ht]
    \centering
   \includegraphics[width=\linewidth]{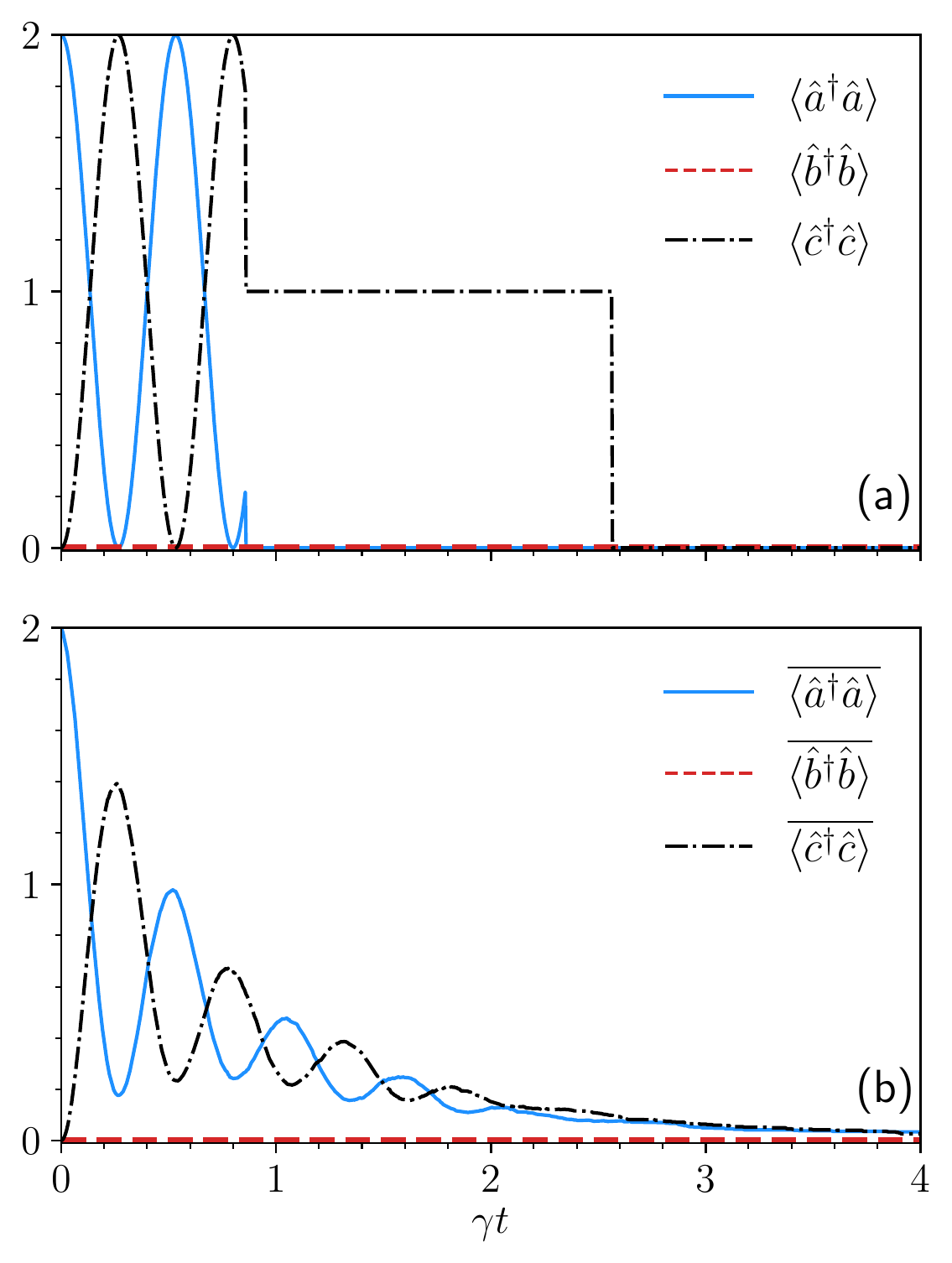}
    \caption{\footnotesize Panel (a) shows an example of a single quantum trajectory, numerically obtained by studying the open quantum dynamics.
    It shows the time evolution of the mean photon number of the left cavity  $\langle \hat{a}^{\dagger} \hat{a} \rangle$ (blue curve), right cavity $\langle \hat{c}^{\dagger} \hat{c} \rangle$ (black dashdotted curve) and of the phonon number of the movable mirror $\langle \hat{b}^{\dagger} \hat{b} \rangle$ (red dashed curve). The system is initialized in $\ket{2,0,0}$ at the resonant condition $\omega_c=\omega_a$, and $\omega_b=3\omega_a/4 $. The numerical simulation initially displays the oscillation predicted by \eqref{evolution_state} until a quantum jump occurs in the right cavity.
    The measure collapse the state into  $-i\ket{0,0,1}$. Even though the two cavities are in resonance, the state $\ket{0,0,1}$ is locked: the photon remains confined in the right cavity. This is an optomechanical feature of our system.% The absence of linear interaction terms denies a one-to-one conversion among the subsystems, as it can be read from the effective Hamiltonian in \eqref{Hopping_Second-order}. 
    After the second jump occurs, the system reaches the state  $\ket{0,0,0}$.
    In panel (b) an average over $500$ trajectories is shown. Clearly, there is a coherent evolution of two photon-pairs state. Such results can be attained as well with a master equation approach, but the locking feature is lost in the average. 
	In both panels, the parameters are $g = 0.06\,\omega_b$, $\omega_a=\omega_c=4\omega_b/3$, and  $\gamma_a = \gamma_b =\gamma_c=\gamma=10^{-4} \omega_b$.}
    \label{Ent_traj}
\end{figure}

\noindent \textbf{Numerical results.}
Figure~\ref{Ent_traj}{\textcolor{blue}{(a)}} shows an example of a single quantum trajectory,  obtained by solving numerically the stochastic evolution of the system wave function. It shows the time evolution of the mean photon number $\langle \hat{a}^{\dagger} \hat{a} \rangle$ (blue curve), $\langle \hat{c}^{\dagger} \hat{c} \rangle$ (black dashdotted curve), of the left and right cavity respectively, and the phonon number $\langle \hat{b}^{\dagger} \hat{b} \rangle$ (red dashed curve). The system is initialized in the state $\ket{2,0,0}$, as in the analytical case. Before a quantum jump occurs, the numerical simulation displays the oscillation predicted by \eqref{evolution_Mean_photon_number}. When the right detector clicks, one photon has escaped from the right cavity. Therefore, the state in \eqref{evolution_state} collapses to $-i\ket{0,0,1}= \hat{c}\ket{\psi (t)}/ [\bra {\psi (t)} \hat{c}^{\dag}\hat{c} \ket{\psi (t)}]^{1/2}$. This state is preserved until a second jump occurs, i.e., the photon remains locked in the right cavity. This is an optomechanical feature of our system. Indeed, the absence of linear interaction terms in \eqref{Hopping_Second-order} denies a one-to-one conversion among the subsystems. Hence, when the second photon jump occurs, it is certain that the state collapses to $\ket{0,0,0}= \hat{c}\ket{0,0,1}$. 

In \figpanel{Ent_traj}{b} the dynamics is shown averaged over $500$ trajectories. Clearly, we see a coherent oscillation of a photon-pair. Of course, in presence of decoherence, such result can be obtained also adopting a master equation approach, but the locking feature emerges only under a post-selection procedure or by studying a single quantum trajectory \cite{Minganti2021a,MacrPRA2022}. Note that, with the parameters used we obtain an effective coupling $\tilde{g}\approx 3\times 10^{-4} \omega_b$, which is almost three times greater than the loss rate $\gamma$ (the latter related to the cavity quality factor $Q$). This regime, defined as strong coupling, allows the photon pairs to flow from one cavity to the other for a certain time before one photon is lost to the environment.

\begin{figure}[t]
    \centering
    \includegraphics[width = \linewidth]{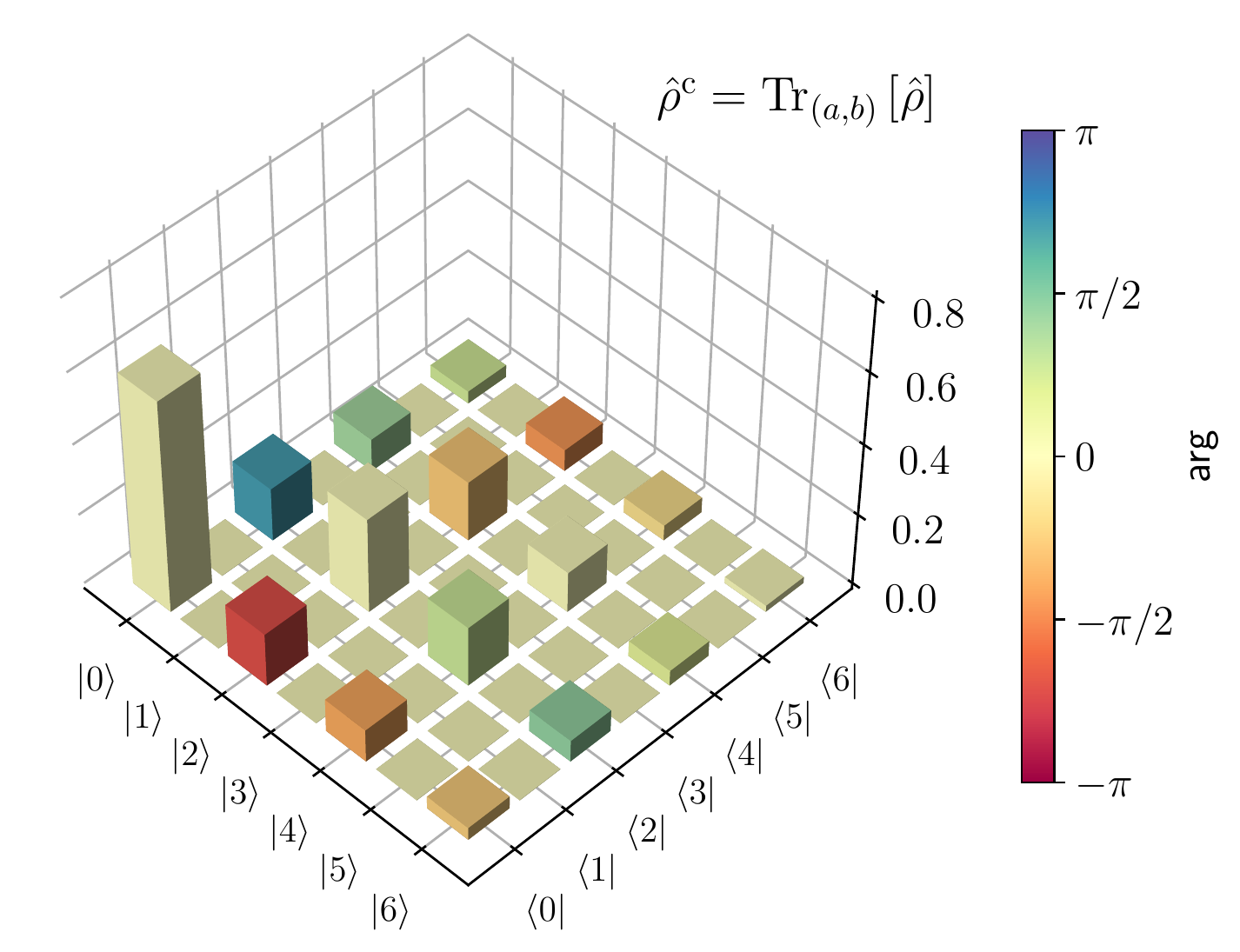}
    \caption{Density matrix elements of the right cavity.
    It is obtained partially tracing over the left cavity and the mirror. Only \textit{even number states} are filled when the right cavity is initially empty and a coherent incoming pulse enters the left cavity. This is in full agreement with the hopping mechanism we proposed. 
    %Inset: the Wigner function representation, exhibits regions of negative probabilities, exemplifying the nonclassical nature of the state of the right cavity. 
    The parameters used here are $g = 0.09\,\omega_b$, $\omega_a=\omega_c=1.1\omega_b$, and  $\gamma_a = \gamma_b =\gamma_c=0$.}
    \label{fig: populations_pulse}
\end{figure}

We conclude this work
considering the case of an incoming Gaussian coherent pulse driving the left cavity while the system is initially in its ground state. For simplicity we present a numerical simulation for the closed dynamics.
%We conclude this work showing that this system can be used as a light filter.
%this system converts coherent incoming light into non-classical squeezed one. 
%To simplify the analysis we perform a numerical simulation for a closed dynamics by taking the right cavity empty and sending a Gaussian coherent pulse to the left cavity.
Figure~\ref{fig: populations_pulse} shows the first matrix elements of the density operator at the end of the dynamics. The state of the right cavity contains only even occupation numbers: in a closed dynamics no loss is possible and the hopping mechanism always involves photon pairs.
%, (ii) the quasi-probability distribution, represented by the Wigner function in the inset panel, has some compact regions of negative probabilities. Moreover, the shape of the  Wigner function clearly shows the presence of quadrature squeezing.

%With the appropriate choice of parameters this converting property from coherent to non-classical squeezed light is scalable, thus we envision its applicability in the context of optomechanical lattices.
% \vspace{4mm}
% \noindent \textbf{Discussion}\\
\section*{Discussion}

\noindent We have %performed a thorough theoretical analysis of
carried out a theoretical analysis of an optomechanical system consisting of two electromagnetic resonators separated by a vibrating two-sided perfect mirror. The Hamiltonian of the system is obtained starting from its canonical quantisation, as shown in Methods, and it accounts also for generic equilibrium positions of the mirror. Our main result %shows that a  photon-pair hopping mechanism, mediated by the phonon vacuum field, appears as a coherent deterministic process in a second-order effective dynamics. 
is the discovery of a photon-pair hopping mechanism, in a coherent second-order effective resonant dynamics.

This effect has been described analytically through the generalized James' approach (see Methods) under the condition $\omega_a=\omega_c$. The numerical analysis of the lowest energy levels showed an avoided-level crossing around the resonant condition [see \figpanel{hopping_position}{c}]. This gap is a trademark of the hybridisation of two photon-pair states. We have performed a stochastic evolution of the system wave function (see Methods) in which we witnessed a coherent oscillatory dynamics between the states $\ket{2,0,0}$ and $\ket{0,0,2}$.  %We have also reported on an effective local analytical description by means of the generalised James' approach (see Methods). 
%Under the choice $\omega_b=3\omega_a/4$, the effective system Hamiltonian in  \eqref{Hopping_Second-order} not only provides effective coupling between the two hybridised photon-pair states in agreement with that one calculated by numerical diagonalisation [see \figpanel{hopping_position}{c}], but it also enables to carry out analytical calculation by solving the stochastic evolution of the system's wave function (see Methods). 
%Indeed, by initialising the system in any of the two photon-pair states components, it has been possible to numerically witness and analytically describe a coherent oscillatory dynamics between the two maximally entangled photon-pairs state (the two photon-pair hopping, see \figref{Ent_traj}).

%Figure~\ref{fig: populations_pulse} shows that the system also works as a 
%quantum converter of light, from coherent to non-classical squeezed states.
%Our results have been obtained by taking into account experimentally attainable setups \cite{Connell2010,rouxinol2016}. 

The effects described here could be experimentally reproduced, with the chosen parameters, in circuit-optomechanical systems by using ultra-high-frequency mechanical micro- or nano-resonators in the GHz spectral range; alternatively, using two \emph{LC} circuits bridged by a SQUID.
%is an experimental  setup capable to show the effect.
Moreover, in arrays of non-linearly coupled cavities \cite{Jin2013}, where the photon crystal associated to a periodic modulation of the photon blockade can emerge, the optomechanical system proposed here allows investigating a new mechanism of photon-pair propagation in optomechanical lattices \cite{Chen2014,Schmidt2015}.

\section*{Methods}
\noindent
\textbf{Derivation of the system Hamiltonian.} We begin by considering  two non-interacting electromagnetic cavities separated by a perfect movable mirror. For simplicity, following Ref.~\cite{Law1995}, we conduct our analysis in $1D$ and generalise it to our case.  To set the notation, $\pm I$ denotes the extremes of the cavity, $M$ and $q(t)$ the mass and the position of the movable mirror respectively. The electromagnetic field, in absence of charges, obeys the wave equation; the motion of the movable mirror is influenced by the radiation pressure of the fields in the two cavities [see  \figpanel{hopping_position}{a} in the main text], so that, it satisfies the Newton's equation
\begin{equation}\label{system}
    \begin{cases}
        \Delta A = 0  \qquad \qquad  x \in \, (-I,q) \, \cup \, (q,I)  \\
        \;M \ddot{q} = -\partial_q V + \frac12 \left[(\partial_- A)^2 - (\partial_+ A)^2 \right]\rvert_{q} 
    \end{cases}
\end{equation}
where $\Delta := \partial_t^2-\partial_x^2$ and $\partial_-, \partial_+$ are the left and right derivatives. 
The potential $V(q)$ is designed  to have infinite walls at the two mirror positions $\pm I$. 
The two radiation pressures $(\partial_\pm A)^2/2$ come with opposite signs and in the form of lateral derivatives, because of the negligible thickness of the movable mirror. 
\smallbreak
By defining $L_k$ and $R_k$ as the Fourier components on the left and right cavity, respectively, the completeness of the mode functions enables to write  
\begin{equation}
    A(t,x) = 
        \begin{cases}
            L^k(t) \, \varphi_k (t,x)  \qquad x \in \, (-I,q)\\
            R^k(t) \, \phi_k (t,x) \qquad x \in \, (q,I)
        \end{cases}
\end{equation}
where the summation in $k$ is understood and 
\begin{align}
    \varphi_k &= \sqrt{\frac{2}{q+I}} \sin{[\omega_k(x+I)]}\;, \nonumber \\
    \phi_k &= \sqrt{\frac{2}{I-q}} \sin{[\Omega_k(x-I)]}\;,
\end{align}
with $\omega_k = k\pi / (q+I) $,  $\Omega_k = k\pi/ (I - q)$. We can still fix a normalisation for $\varphi_k$ and $\phi_k$ choosing
\begin{equation}
    \delta_{ij} = \int_{-I}^q \varphi_i \varphi_j = \int_q^I \phi_i \phi_j\;,
\end{equation}
as the Kronecker delta.
The wave equation~\eqref{system} can be projected along one Fourier component, and the equation of motion of the movable mirror becomes
\bea
\ddot{L}_k + \omega_k^2 L_k &-& \frac{g_{km}\left( 2\dot{q}\dot{L}^m +\ddot{q}L^m \right) }{I+q} \\
&&+ \dot{q}^2 \frac{\left( g_{km} +g_{kj}g^j_{\; \; m} \right) L^m }{(I+q)^2} = 0\; ,\nonumber\\
\ddot{R}_k + \Omega_k^2 R_k &-& \frac{\gamma_{km}\left( 2\dot{q}\dot{R}^m +\ddot{q}R^m \right) }{I-q} \nonumber \\
&&- \dot{q}^2 \frac{\left( \gamma_{km} -\gamma_{kj}\gamma^j_{\; \; m} \right) R^m }{(I-q)^2}  = 0 \nonumber \; , \\
\hspace{-1cm}M \ddot{q} + \partial_q V &+& (-1)^{k+m} \left( \frac{\Omega_k \Omega_m R^k R^m }{I-q} - \frac{\omega_k \omega_m L^k L^m }{q+I}  \right) = 0 \; , \nonumber
\eea
with 
\bea
g_{km} &=& (q+I)\int_{-I}^q \partial_q (\varphi_k)  \varphi_m \nonumber \\
&=& -
\gamma_{km} = -(I-q)\int_q^I \partial_q (\phi_k)   \phi_m \; ,
\eea
that satisfy
\bea
g_{kj}g^j_{\;m} &=& - (q+I)^2 \int_{-I}^q \partial_q \varphi_k \partial_q \varphi_m \nonumber \\
&=& 
\gamma_{kj}\gamma^j_{\;m} = - (I-q)^2 \int_q^I \partial_q \phi_k \partial_q \phi_m \; ,
\eea
and
\begin{align}
    \int_{-I}^q \varphi_k \partial_q^2 \varphi_m &=\frac{1}{(q+I)^2} \left(g_{kj}g^j_{\;\;m} - g_{km} \right)\nonumber \,,\\
    \int_q^I \phi_k \partial_q^2 \phi_m &= \frac{1}{(I-q)^2} \left(\gamma_{kj}\gamma^j_{\;\;m} + \gamma_{km}  \right)\,.  
\end{align}
The system of equations \eqref{system} can be derived from the following Lagrangian
\bea
\mathcal{L}(q,\dot{q}, &&L, \dot{L}, R, \dot{R}) =  \\
&& \frac12 \left( \dot{L}_k \dot{L}^k - \omega^2_k L_k  L^k + \dot{R}_k \dot{R}^k - \Omega^2_k R_k R^k \right) \nonumber \\
&&+ \frac12 M \dot{q}^2 - V - \dot{q} \left( g_{km} \frac{\dot{L}^k L^m}{q+I} +  \gamma_{km} \frac{\dot{R}^k R^m}{I-q} \right) \nonumber \\
&&-
\frac{\dot{q}^2}{2} \left [ g_{kj}g^j_{\;\; m} \frac{L^k L^m}{(q+I)^2} +\gamma_{kj}\gamma^j_{\;\; m} \frac{R^k R^m}{(I-q)^2} \right] \, , \nonumber
\eea
and the corresponding Hamiltonian is
\bea
\label{ClassicalHamiltonian}
\mathcal{H}(q,p,&&L,\Lambda, R, W) = \\
&&\frac{1}{2M}\left(p+ g_{km}\frac{\Lambda^k L^m}{q+I} +  \gamma_{km} \frac{W^k R^m}{I-q} \right)^2 + V \nonumber\\
&&+ \frac12 \left(\Lambda_k \Lambda^k + \omega^2_k L^k L^k \right) + \frac12 \left(W_k W^k + \Omega^2_k R^k R^k \right) \, . \nonumber
\eea
To quantise the Hamiltonian, consider the operators $\{\hat{q}, \hat{p},\hat{L}_k, \hat{\Lambda}_k, \hat{R}_k, \hat{W}_k \}$ and impose the commutation relations ($\hbar=1$), $[\hat{q},\hat{p}] = i$, $ [\hat{L}_k,\hat{\Lambda}_m] = i \delta_{km}$ and $  [\hat{R}_k,\hat{W}_m] = i \delta_{km} $, while $[\hat{q},\hat{L}_m]=[\hat{q},\hat{R}_m]=[\hat{L}_k,\hat{R}_m] = [\hat{p}, \hat{L}_m] =[\hat{p}, \hat{W}_m]= [\hat{\Lambda}_k, \hat{W}_m] = 0$.
Using the ladder operators
\begin{align}\label{a_and_c}
    \hat{a}_k &= \frac{1}{\sqrt{2\omega_k}} \left( \omega_k \hat{L}_k + i \hat{\Lambda}_k \right)\nonumber \;, \\ 
    \hat{c}_k &= \frac{1}{\sqrt{2\Omega_k}} \left( \Omega_k \hat{R}_k + i \hat{W}_k \right)\;,
\end{align}
the Hamiltonian \eqref{ClassicalHamiltonian} becomes

\bea
\label{QuantumHamiltonian}
\hat{H}' &=& \frac{( \hat{p} + \hat{\Gamma} \,)^2}{2M} + \hat{V} +\sum_k \omega_k \hat{a}^\dagger_k \hat{a}_k \nonumber \\
&&+ \sum_k \Omega_k \hat{c}^\dagger_k \hat{c}_k - \frac{\pi q}{6(q+I)(q-I)} \mathds{1}\;,
\eea
where we have already resummed the vacuum point fluctuations, and 
\bea
\hat{\Gamma} &=& \frac{i}{2} \sqrt{\frac{m}{k}} \left[ \frac{g^{km} (\hat{a}^\dagger_k - \hat{a}_k ) (\hat{a}^\dagger_m + \hat{a}_m )}{q+I} \right. \nonumber \\
&&+ \left. \frac{\gamma^{km} (\hat{c}^\dagger_k - \hat{c}_k ) (\hat{c}^\dagger_m + \hat{c}_m )}{I-q} \right]\;.
\eea
This is the full Hamiltonian of the problem. In order to derive \eqref{Hamiltonian} we still need to linearise it and consider the unimodal case. To linearise, first consider $\Gamma \approx \Gamma_0$ constant and then introduce a variation from the expected position of the mirror $q = q_0 + \delta q$, and expand all the terms accordingly 
\bea
\omega_k &=& \frac{k\pi}{q_0+I} \left(1 - \frac{\delta q}{q_0+I}\right) +\mathcal{O}\left[\frac{\delta q^2}{(q_0+I)^2}\right]\;, \nonumber \\
\Omega_k &=& \frac{k\pi}{I-q_0} \left(1 + \frac{\delta q}{I-q_0}\right) + \mathcal{O}\left[\frac{\delta q^2}{(I-q_0)^2}\right]\;, 
\eea
which in turn, from \eqref{a_and_c}, induces 
\bea
\hat{a}_k &\approx& (\hat{a}_0)_k - \frac{\delta q}{2(q_0+I)} (\hat{a}^\dagger_0)_k \;, \nonumber \\
\hat{c}_k &\approx& (\hat{c}_0)_k + \frac{\delta q}{2(I-q_0)} (\hat{c}^\dagger_0)_k\, .
\eea
Performing the unitary transformation $\hat{U}=exp(i \delta q  \hat{\Gamma}_0)$ on \eqref{QuantumHamiltonian}, proves that 
\bea
\hat{H} &=& \hat{U} \hat{H}' \hat{U}^\dagger = \frac{\hat{p}^2}{2M} + \hat{\mathcal{V}} \nonumber \\
&&+ \sum_k \left[ (\omega_0)_k (\hat{a}^\dagger_0)_k (\hat{a}_0)_k  +   (\Omega_0)_k (\hat{c}^\dagger_0)_k (\hat{c}_0)_k   \right] \nonumber \\
&&- \delta q (\hat{G}_0+\hat{F}_0)\;,
\eea
where $\hat{\mathcal{V}} = \hat{V} - \pi q \hat{\mathbbm{1}}/6(q+I)(q-I)$
and 
\bea
&& \hat{F}_0 = \nonumber \\
&&\frac{1}{2(q_0+I)} \sum_{k,j} (-1)^{k+j} \sqrt{(\omega_k \omega_j)_0} \; (\hat{a}^\dagger_k + \hat{a}_k)(\hat{a}^\dagger_m + \hat{a}_m)\;, \nonumber  \\
&&\hat{G}_0 = \nonumber \\
&&-\frac{1}{2(I-q_0)} \sum_{k,j} (-1)^{k+j} \sqrt{(\Omega_k \Omega_j)_0} \; (\hat{c}^\dagger_k + \hat{c}_k)(\hat{c}^\dagger_m + \hat{c}_m)   \;. \nonumber \\
\eea
To finally obtain \eqref{Hamiltonian} in the main text, we consider a quadratic potential $V$ and  introduce the vibrating mirror ladder operators  $\{ b,b^\dagger \}$ in a way that $\delta q= x_{\rm zpf}(b+b^\dagger)$, where  $x_{\rm zpf}$ is the zero-point-fluctuation amplitude of the vibrating mirror. By reducing all the modes to one ($k=j=1$), the system Hamiltonian in \eqref{QuantumHamiltonian} can be written down as
\bea
\hat{H} &=& \omega_a \hat{a}^\dagger \hat{a} + \omega_b \hat{b}^\dagger \hat{b} + \omega_c \hat{c}^\dagger \hat{c} \\
&&+ \frac{x_{\rm zpf}}{2\pi}\left[ \omega_c^2 (\hat{c}+\hat{c}^\dagger)^2- \omega_a^2 (\hat{a}+\hat{a}^\dagger)^2  \right](\hat{b}+\hat{b}^\dagger)\;. \nonumber
\eea
Defining a coupling strength $g=\omega_c^2 \, x_{\rm zpf}/\pi = \omega_c \, x_{\rm zpf} / (I-q_0)$ the \eqref{Hamiltonian} in the main text is obtained. Note that since $\hbar =1$ the coupling strength $g$ has the right units.\\

% \label{james_eff_hamiltonian}
\noindent \textbf{Derivation of the effective Hamiltonians: Applying the generalized James' method.} For interacting quantum systems that are strongly detuned, an effective Hamiltonian can be derived  using the generalized James' effective Hamiltonian method~\cite{Shao2017}. 
To apply this method to \eqref{Hamiltonian}, we first rewrite it in the interaction picture,
\bea
\label{Interaction_Picture}
\hat H_{\rm I}(t) &=& g\left [\hat{c}^{\dag} \hat{c} - \frac{\omega_a^2}{\omega_c^2} \hat{a}^{\dag} \hat{a} \right ] \hat{b} \, e^{-i \omega_b t} \\ 
&&+ \frac{g}{2} \left[ (\hat{c})^2 \hat{b} \, e^{-i (\omega_b+2\omega_c) t}  - \frac{\omega_a^2}{\omega_c^2} (\hat{a})^2 \hat{b} \,e^{-i (\omega_b+2\omega_a) t } \right ]\nonumber\\
&&+ \frac{g}{2} \left[ (\hat{c}^{\dag})^2 \hat{b}\, e^{i (2\omega_c-\omega_b) t}  - \frac{\omega_a^2}{\omega_c^2} (\hat{a}^{\dag})^2 \hat{b}\, e^{i (2\omega_a-\omega_b) t } \right] \nonumber \\
&&+ \rm h.c. \nonumber \, .
\eea
%Depending on the process we want to describe, \eqref{Interaction_Picture} can be expressed in terms of some $\hat{h}_k e^{i \omega_k}$ operators. The oscillation frequencies $\omega_k$  are a combination of the bare transition frequencies, such that, they satisfie  the resonance condition related to the avoided level crossing around which the coherent process takes place. In terms of the $\hat{h}_k e^{i \omega_k}$ the system Hamiltonian in \eqref{Interaction_Picture} can be written as
This can be rewritten as 
\be \label{APP:Heff0} 
\hat H_{\rm I}(t) = \sum_{k} \mleft[ \hat h_k e^{-i \omega_k t} + \hat h_k^\dag e^{ i \omega_k t} \mright] \, .
\ee
where now the $\omega_k$ are a combination of the bare transition frequencies.
It turns out that, a photon-pairs hopping mechanism already appears with a second-order generalized James' effective Hamiltonian method~\cite{Shao2017}. This accounts for calculating  
\be \label{APP:second-oredr}
\hat H_{\rm I}^{(2)} (t) = \sum_{j,k} \frac{1}{\omega_k} \mleft[ \hat h_j \hat h_k^\dag e^{i ( \omega_k -\omega_j) t} - \hat h_j^\dag \hat h_k e^{ i (\omega_j - \omega_k) t} \mright] \, .
\ee
In the rotating-wave approximation, all frequency contributions which are different from zero can be neglected.
Since the frequencies $\omega_k$ are all different, we only keep the terms in $\hat H_{\rm I}^{(2)} (t)$ where the sum of the exponent is zero.  

Starting from \eqref{Interaction_Picture} and considering the resonant condition $\omega_a=\omega_c$, only three terms need to be considered
\begin{align}\label{h_i}
  \hat{h}_1 &= \frac{g}{2} (\hat{c}{^\dagger}^2 - \hat{a}{^\dagger}^2) \hat{b}^\dagger  && \omega_1=2\omega_a+\omega_b\nonumber ,\\
  \hat{h}_2 &= \frac{g}{2} (\hat{c}{^\dagger}^2 - \hat{a}{^\dagger}^2) \hat{b} && \omega_2 = 2\omega_a-\omega_b  \nonumber ,\\
  \hat{h}_3 &= \frac{g}{2} \left( \{ \hat{c},\hat{c}^\dagger  \} - \{ \hat{a},\hat{a}^\dagger  \} \right ) \hat{b}^\dagger && \omega_3 = \omega_b.
\end{align}
From the canonical commutation relations it follows that
\bea
\comm{\hat{h}_1}{\hat{h}_1^\dagger} &=& \frac{g^2}{4} \left [ \hat{a}^2\hat{c}{^\dagger}^2 + \hat{a}{^\dagger}^2\hat{c}^2 -\hat{c}{^\dagger}^2\hat{c}^2 - \hat{a}{^\dagger}^2\hat{a}^2 \right. \nonumber \\
&&+ \left. 2 \hat{b}{^\dagger} \hat{b}   \left( \{ \hat{c},\hat{c}^\dagger  \} + \{ \hat{a},\hat{a}^\dagger  \} \right ) \vphantom{{c^\dagger}^2} \right ] \, , \nonumber \\
\comm{\hat{h}_2}{\hat{h}_2^\dagger} &=& \frac{g^2}{4} \left [ \hat{c}{^\dagger}^2\hat{c}^2 + \hat{a}{^\dagger}^2\hat{a}^2 - \hat{c}{^\dagger}^2\hat{a}^2 - \hat{a}{^\dagger}^2\hat{c}^2 \right. \nonumber \\
&&+ \left. 2 \hat{b} \hat{b}^\dagger   \left( \{ \hat{c},\hat{c}^\dagger  \} + \{ \hat{a},\hat{a}^\dagger  \} \right ) \vphantom{{c^\dagger}^2} \right ] \nonumber, \\
\comm{\hat{h}_3}{\hat{h}_3^\dagger} &=& -g^2 (\hat{c}^\dagger \hat{c} - \hat{a}^\dagger \hat{a})^2 \nonumber ,
\eea
so James' effective Hamiltonian is
\begin{align}
    \hat{H}^{(2)}_{\rm eff} =& \left[ \omega_a + \frac{g^2(4\omega_a+\omega_b)}{8\omega_a^2-2 \omega^2_b}\right](\hat{c}^\dagger \hat{c} + \hat{a}^\dagger \hat{a}) \nonumber \\
    &+  \frac{g^2 (3\omega_b^2- 8 \omega^2_a)}{(8\omega_a^2-2\omega_b^2)\omega_b} [(\hat{c}^\dagger \hat{c})^2 + (\hat{a}^\dagger \hat{a})^2] \nonumber\\
    &+ \left[\omega_b + \frac{ 4 g^2\omega_a}{4\omega_a^2-\omega^2_b} (\hat{c}^\dagger \hat{c} + \hat{a}^\dagger \hat{a} + \mathds{1})\right] \hat{b}^\dagger \hat{b} \nonumber\\
    &+\frac{2g^2}{\omega_b} \hat{a}^\dagger \hat{a} \hat{c}^\dagger \hat{c} + \frac{g^2}{2\omega_a-\omega_b}\mathds{1}\nonumber \\
    &-\frac{g^2\omega_b}{8\omega_a^2-2\omega^2_b} (\hat{a}^2 \hat{c}{^\dagger}^2 + \hat{a}{^\dagger}^2 \hat{c}^2 ) .
\end{align}
All the terms but the last one are energy shifts. The latter is the desired hopping mechanism.\\

% \label{MCWF_Method}

\noindent \textbf{Monte Carlo wave function approach: Quantum trajectory.} Following Refs.~\cite{Dalibard1992,Molmer1993}, in order to describe the Monte Carlo wave function (MCWF) approach, we introduce the non-Hermitian Hamiltonian
\be \label{non_Hermitian_H} 
\mathcal{\hat H} = \hat H - \frac{i}{2} \sum_m \gamma_m \, \hat \Gamma_m^{\dag} \hat \Gamma_m \, ,
\ee
describing the effect of the environment between two quantum jumps. 
Here, $\hat{H}$ represents the Hamiltonian part of the dynamics, and one can either use the full or the effective Hamiltonian, while $\hat \Gamma_m$ are the jump operators.
The evolution of a quantum trajectory is thus dictated by a non-Hermitian evolution via $\mathcal{\hat H}$ interrupted by random quantum jumps. The algorithm to obtain such a dynamics reads:

\noindent (i)\,\, $\ket{\psi (t)}$ is the normalized wave function at the initial time $t$.

\noindent (ii)\,\, The probability that a quantum jump occurs through the $m$-th dissipative channel in a small amount of time $dt$ is
\be
\delta p_m(t) =dt \gamma_m \, 
\langle\psi ( t)|\hat \Gamma_m^{\dag} \hat \Gamma_m |\psi (t)\rangle,
\ee
such that $\delta p_m(t)\ll 1$.

\noindent (iii)\,\,  One randomly generates a real number $\varepsilon \in [0,1]$. 

\noindent (iv)\,\,  If $\sum_m \delta p_m(t)<\varepsilon$, no quantum jump occurs, and the system evolves as
\be \label{evolution} 
\ket{\psi (t+ dt)} = \exp \mleft( - i \mathcal{\hat H} dt \mright) =  \mathbbm{1}- i dt \mathcal{\hat H} \ket{\psi (t)} + \mathcal{O}(dt^2) \, .
\ee

\noindent (v)\,\,  Otherwise, if $\sum_m \delta p_m(t)>\varepsilon$, a quantum jump occurs. To decide which channel dissipates, a second random number $\varepsilon'$ is generated, and each quantum jump is selected with probability $\delta p_m(t)/(\sum_n \delta p_n(t))$.
The wave function then becomes
\be \label{jump} 
\ket{\psi (t + dt)} =  \hat \Gamma_m \ket{\psi (t)}.
\ee

\noindent (vi)\,\,  At this point, independently of whether a quantum jump took place, the wave function $\ket{\psi (t + dt)}$ is renormalized and used for the next step of the time evolution.
\vspace{0.2cm}

Any quantum jump corresponds to the projection of the wave function associated with a generalized measurement process (wave-function collapse through a positive operator-valued measure) \cite{Wiseman_BOOK_Quantum}. 
Although the results of MCWF recovers those of the Lindblad master equation, by averaging over an infinite number of trajectories, noise effects determine the convergence rate. 

%Before proceeding further, let us clarify the notation. In the following, By odopting the notation $\langle \hat{O}(t)\rangle$ we mean the expecation value of the operator $\hat{O}$ on a single quantum trajectory, i.e., $\langle \hat{O}(t)\rangle\equiv \langle \psi(t)|\hat{O}|\psi(t) \rangle$, while we indicate average quantities obtained over an ideally infinite number of quantum trajectories  as $ \overline{\langle \hat{O}(t) \rangle} = \overline{\langle \psi(t)|\hat{O}|\psi(t) \rangle}$. 

\bibliography{Riken}% Produces the bibliography via BibTeX.
\section*{}
\noindent \textbf{Acknowledgements}

\noindent F.N. is supported in part by: Nippon Telegraph and Telephone Corporation (NTT) Research, the Japan Science and Technology Agency (JST) [via the Quantum Leap Flagship Program (Q-LEAP), and the Moonshot R$\&$D Grant Number JPMJMS2061], the Japan Society for the Promotion of Science (JSPS) [via the Grants-in-Aid for Scientific Research (KAKENHI) Grant No. JP20H00134], the Army Research Office (ARO) (Grant No. W911NF-18-1-0358), the Asian Office of Aerospace Research and Development (AOARD) (via Grant No. FA2386-20-1-4069), and the Foundational Questions Institute Fund (FQXi) via Grant No. FQXi-IAF19-06.
S.S. acknowledges the Army Research Office (ARO) (Grant No. W911NF-19-1-0065). R.L.F. is supported by European Union (Next Generation EU) via MUR D.M. Grant No. 737-2021. 

\end{document}